\documentclass[aps,pra,twocolumn,showpacs]{revtex4}
\usepackage[tbtags]{amsmath}
\usepackage{amssymb,amsfonts,bm,mathptmx,enumerate}

\renewcommand{\vec}[1]{\mathbf{#1}}

\renewcommand{\geq}{\geqslant}
\DeclareSymbolFont{symbols}{OMS}{cmsy}{m}{n}

\begin{document}

\title{Einstein-Podolsky-Rosen correlations and Galilean
  transformations}

\author{Pawe{\l} Caban}
\email{P.Caban@merlin.fic.uni.lodz.pl}
\author{Jakub Rembieli{\'n}ski}
\email{J.Rembielinski@merlin.fic.uni.lodz.pl}
\author{Kordian A.\ Smoli{\'n}ski}
\email{K.A.Smolinski@merlin.fic.uni.lodz.pl}
\author{Zbigniew Walczak}
\email{Z.Walczak@merlin.fic.uni.lodz.pl}

\affiliation{Department of Theoretical Physics, University of
  {\L}{\'o}d{\'z}, ul.\ Pomorska 149/153, 90-236 {\L}{\'o}d{\'z}, Poland}

\date{30 September 2002}

\begin{abstract}
  In this paper we calculate with full details Einstein-Podolsky-Rosen
  spin correlations in the framework of nonrelativistic quantum
  mechanics. We consider the following situation: two-particle state
  is prepared (we consider separately distinguishable and identical
  particles and take into account the space part of the wave function)
  and two observers in relative motion measure the spin component of
  the particle along given directions. The measurements are performed
  in bounded regions of space (detectors), not necessarily
  simultaneously.  The resulting correlation function depends not only
  on the directions of spin measurements but also on the relative
  velocity of the observers.
\end{abstract}

\pacs{03.65.Ud}

\maketitle

\section{Introduction}
\label{cab_korel_sec_intro}

The issue of locality plays a central role in recent theoretical and
experimental investigations of basic properties of quantum mechanics.
The history of this long-standing problem began in 1935, when
Einstein, Podolsky, and Rosen (EPR) published their paper
\cite{cab_korel_EPR1935}.  EPR considered a gedanken experiment with
two spatially separated particles, $a$ and $b$, in an entangled state
in which the relative position, $x_a-x_b$, and the total momentum,
$p_a+p_b$, have definite values.  If the momentum of the particle $a$
is measured one can predict with certainty the momentum of the
particle $b$.  Since particles are spatially separated and the
locality of quantum mechanics is assumed, the measurement on the
particle $a$ does not disturb the particle $b$, thus, due to the EPR
reality criterion, the momentum of the particle $b$ is an element of
reality.  Alternatively one could measure the position of the particle
$a$, and by the same arguments one concludes that also the position of
the particle $b$ is an element of reality.  But quantum mechanics does
not allow us to find simultaneously values of $p_b$ and $x_b$,
therefore EPR concluded that the description of reality that is
provided by quantum mechanics is not complete.

In the above so-called ``EPR paradox,'' the Einstein locality
principle, which states ``\ldots the real factual situation of the system
S2 is independent on what is done with the system S1, which is
spatially separated from the former,'' \cite{cab_korel_Einstein1949}
was applied for the first time explicitly to quantum mechanics.

For a long time, EPR predictions were experimentaly untestable.  The
problem was reformulated in terms of spin variables by Bohm
\cite{cab_korel_Bohm1951}, and in 1964 Bell proved that
\cite{cab_korel_Bell1964} in such a setting some inequality should
hold for any local realistic theory.  The Bell inequality was easier
to handle experimentally because it imposes some constraints only on
correlations of results of measurements performed by two distant
observers.  Many experiments were performed to test Bell-type
inequalities
\cite{cab_korel_Aspect1981,cab_korel_Aspect1982,%
cab_korel_Tittel1997,cab_korel_Tittel1998,cab_korel_Weihs1998,%
cab_korel_Apostolakis1998,cab_korel_Foadi1999}, %
and all of them showed that they are violated and that
quantum-mechanical predictions are satisfied \footnote{For technical
  reasons, there may exist some loopholes (see, e.g.,
  \cite{cab_korel_Gisin1999} and references therein) that allow us to
  maintain the local and realistic interpretation of quantum
  mechanics, but we will not consider them here.}.  Recently, even
experiments with observers in relative motion were performed
\cite{cab_korel_Zbinden2000}.

In the standard formulation of the Bell inequality, only the spin part
of the wave function of two particles is taken into account
\cite{cab_korel_Ballentine1998,cab_korel_Peres1995}.  However, some
authors pointed out that when the issue of locality in quantum
mechanics is considered, the space part of the wave function cannot be
neglected \cite{cab_korel_Volovich2000,cab_korel_Volovich2001}.  We
accept this point of view.  Unfortunately, in the standard formulation
of relativistic quantum mechanics, the notion of localization of a
particle is ill-defined.  The main problem one encounters in this case
concerns the Lorentz covariance of the localization (see, e.g.,
\cite{cab_korel_Bacry1988}).  It causes the framework of the standard
relativistic quantum mechanics to be unsuitable for the calculation of
EPR correlations in the most general case, i.e., when the space part
of the wave function and the relative motion of the observers are
taken into account.  In the framework of standard relativistic quantum
mechanics, the spin correlations were calculated in
\cite{cab_korel_Czachor1997_1,cab_korel_Czachor1997_2} but the
derivation of the correlation function presented therein does not
involve localization of measured particles in detectors and is
restricted to the measurements performed in the same inertial frame.
The framework of the Lorentz covariant quantum mechanics developed in
\cite{cab_korel_CR1999} seems to be more suitable to calculate the EPR
correlation function in the general case.  The correlation function in
such a framework was calculated in \cite{cab_korel_RS2002}.

On the other hand, to the best of our knowledge,no systematic review
of EPR correlations in the framework of nonrelativistic quantum
mechanics in the general case exist in the literature.  One of the
reasons for this is that the EPR paradox appears only in the
relativistic case.  But in our opinion, we should know also the exact
form of the EPR correlation function in the case of nonrelativistic
quantum mechanics, at least to compare it with results obtained in the
relativistic case.  Therefore, the main goal of our present paper is
to calculate in detail spin correlations in the framework of
nonrelativistic quantum mechanics.  More precisely, we consider the
following situation: a two-particle state is prepared (we consider
separately distinguishable and identical particles and we take into
account the space part of the wave function) and two observers in
relative motion measure the spin component of the particle along given
directions. The measurements are performed in bounded regions of space
(detectors) not necessarily simultaneously.

The paper is organized as follows.  In
Sec.~\ref{cab_korel_sec_correlations_without_destruction}, we
calculate EPR spin correlations taking into account the space part of
the wave function and motion of the observers.  We discuss the cases
of distinguishable and identical particles separately.
Section~\ref{cab_korel_sec_conclusions} concludes with a summary of
our results.  The main facts concerning the Galilean group and its
unitary ray representations are collected in the Appendix.

\section{\label{cab_korel_sec_correlations_without_destruction} EPR
  correlations}

In this section we calculate quantum correlations in the following
case.  In a given inertial frame of reference $\mathcal{O}$, a
two-particle state is prepared.  Two observers, say $\mathcal{A}$ and
$\mathcal{B}$, travel with constant velocities with respect to the
frame $\mathcal{O}$.  Each observer possesses a detector which can
measure the spin component of a particle along a given axis fixed by
unit vectors $\vec{a}$ and $\vec{b}$, respectively.  We assume that
the spin measurements take place only if the particle is inside the
detector.  Thus we assume that detectors occupy regions $A$ and $B$,
respectively.  We consider separately the case of distinguishable and
identical particles.  For the notation concerning the Galilean group
and its unitary ray representations, see the Appendix.

\subsection{\label{cab_korel_subsection_without_destruct_distinguishable}
  Distinguishable particles}

We consider two spin $s$ particles, say $\alpha$ and $\beta$.  We assume that
spins of both particles are equal for simplicity, however it is
straightforward to generalize our considerations for the case of
particles with different spins.  The space of states of this
two-particle system is $\mathcal{H}^\alpha \otimes \mathcal{H}^\beta$, where
$\mathcal{H}^\alpha$ and $\mathcal{H}^\beta$ denote the space of states of the
particles $\alpha$ and $\beta$, respectively.  In the spaces ${\mathcal{H}}^\alpha$
and ${\mathcal{H}}^\beta$, we will use bases $\{|\vec{x}_\alpha,\vec{n}_\alpha,\lambda_\alpha\rangle\}$
and $\{|\vec{x}_\beta,\vec{n}_\beta,\lambda_\beta\rangle\}$, respectively.  A vector
$|\vec{x}_\alpha,\vec{n}_\alpha,\lambda_\alpha\rangle$ ($|\vec{x}_\beta,\vec{n}_\beta,\lambda_\beta\rangle$) describes
the situation in which the particle $\alpha$ ($\beta$) is localized at
$\Vec{x}_\alpha$ ($\vec{x}_\beta$) and its spin component along the direction
determined by a unit vector $\vec{n}_\alpha$ ($\vec{n}_\beta$) is equal to
$\lambda_\alpha$ ($\lambda_\beta$).  Definition of the vectors $|\Vec{x},\Vec{n},\lambda\rangle$ and
their basic properties are given in the Appendix,
Eqs.~\eqref{cab_korel_vectors_xnlambda_eq1}
and~\eqref{cab_korel_vectors_xnlambda_eq2}.  We want to describe an
EPR-type experiment in which two distant observers ${\mathcal{A}}$ and
${\mathcal{B}}$ measure spin components of the particles using
detectors that occupy some bounded regions $A$ and $B$, respectively.
Thus the measurement consists of the localization inside the region of
the detector and simultaneous measurement of the spin component.
Therefore, corresponding observables for particles $\alpha$ and $\beta$ read
\begin{equation}
  \Lambda_{A,\vec{a}}^{s}\otimes I,\qquad 
  I\otimes\Lambda_{B,\vec{b}}^{s},
  \label{cab_korel_Lambda_a_b_two}
\end{equation}
where the spectral decomposition of $\Lambda_{\Omega,\vec{n}}^{s}$ is the
following:
\begin{equation}
  \Lambda_{\Omega,\vec{n}}^{s}  =
  \sum_{\lambda=-s}^{s}\lambda\left(\int_{\Omega}d^3 \vec{x}\,
    |\vec{x},\vec{n},\lambda\rangle
    \langle\vec{x},\vec{n},\lambda|\right) 
  \equiv\sum_{\lambda=-s}^{s}\lambda\,\,
  \Pi_{\Omega,\vec{n}}^{s,\lambda}.
  \label{cab_korel_spectral_Lambda_dist}
\end{equation}
The projectors $\Pi_{\Omega,\vec{n}}^{s,\lambda}$ in
Eq.~\eqref{cab_korel_spectral_Lambda_dist} have the following obvious
interpretation: When we measure $\Pi_{\Omega,\vec{n}}^{s,\lambda}$ we get the value
1 if and only if the corresponding particle is inside $\Omega$ and its spin
component along the direction $\Vec{n}$ is equal to $\lambda$.

Under Galilean boosts, projector $\Pi_{\Omega,\vec{n}}^{s,\lambda}$ transforms as
follows [cf.\ Eq.~\eqref{cab_korel_app_action_position_n_eq4}]:
\begin{equation}
  \begin{split}
    U^{\dagger}_{t}(\vec{v})\, \Pi_{\Omega,\vec{n}}^{s,\lambda}\, U_t(\vec{v}) &=  
    \int_\Omega d^3\!\vec{x}\, |\vec{x}-t\vec{v},\vec{n},\lambda\rangle
    \langle\vec{x}-t\vec{v},\vec{n},\lambda|\\
    &= \int_{\Omega^\prime(t)} d^3\!\vec{x}\, |\vec{x},\vec{n},\lambda\rangle
    \langle\vec{x},\vec{n},\lambda|\\
    &= \Pi_{\Omega^\prime(t),\vec{n}}^{s,\lambda}, 
  \end{split}
  \label{cab_korel_U_dagger_v_Pi_U_v}
\end{equation}
where $\Omega^\prime(t)=\{\vec{x}^\prime\colon \vec{x}^\prime= \vec{x}-\vec{v}t,\vec{x} \in\Omega\}$.
This means that localization in nonrelativistic quantum mechanics is
covariant; that is, the projector $\Pi_{\Omega^\prime(t),\Vec{n}}^{s,\lambda}$
corresponds to the localization in the same region as seen by the
moving observer, at the moment $t$.  We point out that this is not
true in standard relativistic quantum mechanics
\cite{cab_korel_Bacry1988}.

Now we can calculate quantum correlations.  This can be done in the
following steps.
\begin{enumerate}[(i)]
\item \textit{Preparation of the initial state}.  We assume that a
  two-particle state $\rho$ is prepared in a certain inertial frame of
  reference ${\mathcal{O}}$.  Two other inertial frames of reference,
  ${\mathcal{A}}$ and ${\mathcal{B}}$, move with constant velocities
  with respect to ${\mathcal{O}}$.  We denote the velocity of the
  frame ${\mathcal{O}}$ with respect to ${\mathcal{A}}$ and
  ${\mathcal{B}}$ by $\Vec{v}_A$ and $\vec{v}_B$, respectively.
\item \textit{Measurement performed by observer ${\mathcal{A}}$}.  An
  observer at rest with respect to ${\mathcal{A}}$ (for simplicity, we
  refer to him as to the observer ${\mathcal{A}}$) measures at time
  $t_A$ the observable $\Lambda_{A,\vec{a}}^{s}\otimes I$.  As a result of the
  measurement with selection, ${\mathcal{A}}$ receives a value $\lambda_\alpha$.
\item \textit{Free time evolution of the state}.  Next, the state
  evolves freely in time from $t_A$ to $t_B\geq t_A$.
\item \textit{Measurement performed by observer ${\mathcal{B}}$}.  At
  time $t_B$, an observer that rests with respect to ${\mathcal{B}}$
  (we call him observer ${\mathcal{B}}$) measures
  $I\otimes\Lambda_{B,\vec{b}}^{s}$.  The result of this measurement with
  selection is $\lambda_\beta$.
\end{enumerate}

Let us denote the probability that ${\mathcal{A}}$ receives $\lambda_\alpha$ and
${\mathcal{B}}$ receives $\lambda_\beta$ as $p(\lambda_\alpha,\lambda_\beta)$.  In the case of
distinguishable particles, we define the following correlation
function:
\begin{equation}
  {\mathcal{C}}^{\alpha,\beta}(\vec{a},\vec{b})=
  \sum_{\lambda_{\alpha},\lambda_{\beta}} \lambda_{\alpha} \lambda_{\beta}\, p(\lambda_{\alpha},\lambda_{\beta}).
  \label{cab_korel_Cab_rozne_no_destruction}
\end{equation}
We could also imagine the situation in which the observers do not
distingiush the type of particles.  In such a case, the correlation
function \eqref{cab_korel_Cab_rozne_no_destruction} should be replaced
by
\begin{equation}
  \begin{split}
    {\mathcal{C}}(\vec{a},\vec{b}) 
    &= \sum_{\lambda_\alpha,\lambda_\beta} \lambda_\alpha \lambda_\beta \left[p(\lambda_{\beta},\lambda_{\alpha}) + p(\lambda_{\alpha},\lambda_{\beta})\right]\\
    &={\mathcal{C}}^{\alpha,\beta}(\vec{a},\vec{b})+
    {\mathcal{C}}^{\beta,\alpha}(\vec{a},\vec{b}).
  \end{split}
\end{equation}

Thus we have to calculate $p(\lambda_\alpha,\lambda_\beta)$.  We do it according to the
steps described above.
\begin{enumerate}[(i)]
\item \textit{Preparation of the initial state}.  An initial state is
  prepared in the frame ${\mathcal{O}}$.  At time, $t_A$ it is given
  by $\rho(t_A)$.
\item \textit{Measurement performed by the observer ${\mathcal{A}}$}.
  For the observer ${\mathcal{A}}$, the density matrix $\rho(t_A)$ has
  the form
  \begin{equation}
    \rho_{\mathcal{A}}(t_A)=
    U_{t_A}(\vec{v}_A)\,\rho(t_A)\,U^{\dagger}_{t_A}(\vec{v}_A), 
    \label{cab_korel_stan_rho_A_od_TA_rozne_bez_anihilacji}
  \end{equation}
  where $U_{t}(\vec{v})=U_{t}^{\alpha}(\vec{v}) \otimes U_{t}^{\beta}(\vec{v})$ and
  the unitary operator of pure Galilean boost $U_{t}^{\alpha}(\vec{v})$
  [$U_{t}^{\beta}(\vec{v})$] is given in the Appendix; see
  Eqs.~\eqref{cab_korel_app_unitary_representants},
  \eqref{cab_korel_app_action_momentum_eq4},
  \eqref{cab_korel_app_action_position_eq4},
  and~\eqref{cab_korel_app_action_position_n_eq4}.  Now the observer
  ${\mathcal{A}}$ measures $\Lambda_{A,\Vec{a}}^{s}\otimes I$ in the state
  \eqref{cab_korel_stan_rho_A_od_TA_rozne_bez_anihilacji} and as a
  result of the measurement with selection he receives $\lambda_\alpha$ with the
  probability
  \begin{equation}
    p(\lambda_\alpha)=\mathrm{Tr}
    \left[\rho_{\mathcal{A}}(t_A)\left(\Pi_{A,\vec{a}}^{s,\lambda_\alpha}
        \otimes I\right)\right].
  \end{equation}
  The measurement reduces the density matrix
  \eqref{cab_korel_stan_rho_A_od_TA_rozne_bez_anihilacji} to
  \begin{equation}
    \rho_{\mathcal{A}}^{\lambda_{\alpha}}(t_A)=\frac{\left(
        \Pi_{A,\Vec{a}}^{s,\lambda_\alpha}\otimes I\right)\rho_{\mathcal{A}}(t_A)
      \left(\Pi_{A,\Vec{a}}^{s,\lambda_\alpha}\otimes I\right)}%
    {\mathrm{Tr}\left[\rho_{\mathcal{A}}(t_A)\left(
          \Pi_{A,\Vec{a}}^{s,\lambda_\alpha} 
          \otimes I\right)\right]}.
    \label{cab_korel_rho_po_redukcji_rozne_bez_destr}
  \end{equation}
\item \textit{Free time evolution of the state}.  The density matrix
  \eqref{cab_korel_rho_po_redukcji_rozne_bez_destr} as seen from the
  frame ${\mathcal{O}}$ reads
  \begin{equation}
    \rho^{\lambda_{\alpha}}(t_A)=
    U^{\dagger}_{t_A}(\Vec{v}_A)\,\rho_{\mathcal{A}}^{\lambda_{\alpha}}(t_A)\, 
    U_{t_A}(\Vec{v}_A). 
  \end{equation}
  Now the state $\rho^{\lambda_\alpha}(t_A)$ evolves from time $t_A$ to $t_B$ and
  the resulting density matrix reads
  \begin{equation}
    \rho^{\lambda_{\alpha}}(t_B)=
    U^\dagger(t_B-t_A)\,\rho^{\lambda_{\alpha}}(t_A)\,U(t_B-t_A),
    \label{cab_korel_stan_rho_alpha_od_TB_rozne_bez_anihilacji}
  \end{equation}
  where $U(t_B-t_A)=U^{\alpha}(t_B-t_A)\otimes U^{\beta}(t_B-t_A)$ and $U(t)$ denotes
  the time evolution operator.
\item \textit{Measurement performed by the observer ${\mathcal{B}}$}.
  The density matrix
  \eqref{cab_korel_stan_rho_alpha_od_TB_rozne_bez_anihilacji} as seen
  by the observer ${\mathcal{B}}$ has the form
  \begin{equation}
    \rho^{\lambda_{\alpha}}_{\mathcal{B}}(t_B)=
    U_{t_B}(\Vec{v}_B)\,\rho^{\lambda_{\alpha}}(t_B)\, 
    U^{\dagger}_{t_B}(\Vec{v}_B). 
    \label{cab_korel_stan_B_tB_rozne_bez_anihilacji} 
  \end{equation}
  Now the observer ${\mathcal{B}}$ measures $I\otimes\Lambda_{B,\Vec{b}}^{s}$ in
  the state \eqref{cab_korel_stan_B_tB_rozne_bez_anihilacji} and
  receives $\lambda_\beta$ with the probability
  \begin{equation}
    p(\lambda_\beta|\lambda_\alpha)=\mathrm{Tr}\left[
      \rho_{\mathcal{B}}^{\lambda_{\alpha}}(t_B)\left( 
        I\otimes\Pi_{B,\Vec{b}}^{s,\lambda_\beta}\right)\right].
    \label{cab_korel_prawdop_warunkowe_rozne_bez_anihilacji} 
  \end{equation}
  It is conditional probability because the state in which
  ${\mathcal{B}}$ performs the measurement has the form
  (\ref{cab_korel_stan_rho_alpha_od_TB_rozne_bez_anihilacji}) only if
  ${\mathcal{A}}$ receives $\lambda_\alpha$ in the first measurement.
\end{enumerate}

So finally we get
\begin{equation}
  p(\lambda_\alpha,\lambda_\beta)=p(\lambda_\alpha)\,
  p(\lambda_\beta|\lambda_\alpha).
\end{equation}
Taking into account
Eqs.~\eqref{cab_korel_stan_rho_A_od_TA_rozne_bez_anihilacji}--%
\eqref{cab_korel_prawdop_warunkowe_rozne_bez_anihilacji}, we can write
\begin{widetext}
  \begin{align}
    p(\lambda_\alpha,\lambda_\beta) = & 
    \mathrm{Tr}\Big\{
    \Big[U^{\dagger}_{t_A}(\Vec{v}_A)\left(
      \Pi_{A,\Vec{a}}^{s,\lambda_\alpha}\otimes I\right)
    U_{t_A}(\Vec{v}_A)\Big]
    \rho(t_A)
    \Big[U^{\dagger}_{t_A}(\Vec{v}_A)\left(
      \Pi_{A,\Vec{a}}^{s,\lambda_\alpha}\otimes I\right)
    U_{t_A}(\Vec{v}_A)\Big] \nonumber\\
    &\times
    \Big[U(t_B-t_A)\, U^{\dagger}_{t_B}(\Vec{v}_B)\left(
      I\otimes\Pi_{B,\Vec{b}}^{s,\lambda_\beta}\right)
    U_{t_B}(\Vec{v}_B)\, U^\dagger(t_B-t_A)\Big] 
    \Big\}.
    \label{cab_korel_prawdop_pary_alpha_beta_rozne_bez_anihilacji}
  \end{align}
  Inserting
  Eq.~\eqref{cab_korel_prawdop_pary_alpha_beta_rozne_bez_anihilacji}
  into Eq.~\eqref{cab_korel_Cab_rozne_no_destruction}, we get
  \begin{equation}
    {\mathcal{C}}^{\alpha\beta}(\Vec{a},\Vec{b}) = \mathrm{Tr}\Big\{ \rho(t_A)
    \Big[U^{\alpha\dagger}_{t_A}(\Vec{v}_A)\, \Lambda_{A,\Vec{a}}^{s}\,
    U_{t_A}^{\alpha}(\Vec{v}_A)\Big] 
    \otimes \Big[U^{\beta}(t_B-t_A)\, U^{\beta\dagger}_{t_B}(\Vec{v}_B)\,
    \Lambda_{B,\Vec{b}}^{s}\, U_{t_B}^{\beta}(\Vec{v}_B)\, U^{\beta\dagger}(t_B-t_A)\Big]
    \Big\}.
 \label{funkcja_korel_a_b_rozne_bez_anihil_rozne_czasy_rozne_uklady_eq_1}
 \end{equation}
 The function ${\mathcal{C}}^{\beta\alpha}(\Vec{a},\Vec{b})$ can be obtained
 from ${\mathcal{C}}^{\alpha\beta}(\Vec{a},\Vec{b})$ by simultaneous change of
 the order in the tensor product and change of indices $\alpha$ and $\beta$.
 
 The formula
 \eqref{funkcja_korel_a_b_rozne_bez_anihil_rozne_czasy_rozne_uklady_eq_1}
 may be simplified in a particular case when the initial state is a
 pure one. In this case $\rho(t_A)=|\psi\rangle\langle\psi|$, where
 $|\psi\rangle\in{\mathcal{H}}^\alpha\otimes{\mathcal{H}}^\beta$ is normalized.
 
 Moreover, one can check that in the free time evolution case, the
 following relation holds:
 \begin{equation}
   U(\tau) [U^{\dagger}_{t}(\Vec{v})\, \Pi_{\Omega,\Vec{n}}^{s,\lambda}\, U_t(\Vec{v})] U^\dagger(\tau)  = 
   \frac{1}{(2\pi)^3}\iint d^3 \Vec{k}\,d^3 \Vec{p} \int_\Omega
   d^3 \Vec{x}\,
   e^{i(\Vec{x}-t\Vec{v}) \cdot (\Vec{p}-\Vec{k})+
     [i\tau(\Vec{k}^2-\Vec{p}^2)/2M]} 
   |\Vec{k},\Vec{n},\lambda\rangle\langle\Vec{p},\Vec{n},\lambda|.
   \label{U_tau_U_dagger_v_Pi_U_v_U_dagger_tau}
 \end{equation}
 Thus inserting Eq.~\eqref{cab_korel_U_dagger_v_Pi_U_v} and
 \eqref{U_tau_U_dagger_v_Pi_U_v_U_dagger_tau} into
 Eq.~\eqref{funkcja_korel_a_b_rozne_bez_anihil_rozne_czasy_rozne_uklady_eq_1},
 we have
 \begin{align}
   {\mathcal{C}}^{\alpha\beta}_{\psi}(\vec{a},\vec{b}) =& 
   \frac{1}{(2\pi)^3} 
   \int_A d^3 \vec{x} \int_B d^3 \vec{y} \iint
   d^3 \vec{k}\,d^3 \vec{p}\Big\{
   e^{i(\vec{y}-\vec{v}_Bt_B)\cdot(\vec{p}-\vec{k})+%
     [i(t_B-t_A)/2M_\beta](\vec{k}^2-\vec{p}^2)}\nonumber\\
   &\times
   \sum_{\lambda_\alpha,\lambda_\beta} \lambda_\alpha
   \lambda_\beta \langle\psi|
   \left[ |\vec{x}-\vec{v}_At_A,\vec{a},\lambda_\alpha\rangle
   \langle\vec{x}-\vec{v}_At_A,\vec{a},\lambda_\alpha| \otimes
   |\vec{k},\vec{b},\lambda_\beta\rangle
   \langle\vec{p},\vec{b},\lambda_\beta| 
   \right]|\psi\rangle\Big\},
   \label{cab_korel_f_korel_a_b_rozne_bez_anihil_stan_czysty}  
 \end{align}
 where $M_\beta$ denotes the mass of the particle $\beta$.  In the position
 representation, $|\psi\rangle$ is of the form
 \begin{equation}
   |\psi\rangle =  
   \sum_{m_\alpha,m_\beta} \iint
   d^3 \vec{x}\,d^3 \vec{y}\,\, \psi_{m_\alpha
     m_\beta}(\vec{x},\vec{y}) |\vec{x},m_\alpha\rangle\otimes
   |\vec{y},m_\beta\rangle.
   \label{cab_korel_funkcja_psi_polozeniowa_rozne}
 \end{equation}
 Note that in Eq.~\eqref{cab_korel_funkcja_psi_polozeniowa_rozne} we
 use the basis $\{|\vec{x},m_\alpha\rangle\otimes|\vec{y},m_\beta\rangle\}$, where $m_\alpha$ and $m_\beta$
 denote spin components along the $z$ axis.  One can check that under
 the definition \eqref{cab_korel_funkcja_psi_polozeniowa_rozne} and
 using Eq.~\eqref{cab_korel_iloczyn_y_m_x_n_lambda}
 and~\eqref{cab_korel_iloczyn_p_m_x_n_lambda} we obtain
 \begin{equation}
   \begin{split}
     {\mathcal{C}}^{\alpha\beta}_{\psi}(\vec{a},\vec{b}) = & \frac{1}{(2\pi)^6} \int_A
     d^3 \vec{x} \int_B d^3 \vec{y} \iint d^3 \vec{k}\, d^3 \vec{p}\,
     e^{i(\vec{y}-\vec{v}_Bt_B)\cdot(\vec{p}-\vec{k})+
       [i(t_B-t_A)/2M_B](\vec{k}^2-\vec{p}^2)}
     \sum_{\substack{m_\alpha,m_\beta\\
         m_{\alpha}^{\prime},m_{\beta}^{\prime}}} \iint
     d^3 \vec{x}^\prime \, d^3 \vec{y}^\prime \,
     e^{i(\vec{k}\cdot\vec{x}^\prime-\vec{p}\cdot\vec{y}^\prime)}
     \\ 
     &\quad \times \psi^{\star}_{m_{\alpha}^{\prime}m_{\beta}^{\prime}} 
     (\vec{x}-\vec{v}_At_A,\vec{x}^\prime) 
     \psi_{m_{\alpha}m_{\beta}}(\vec{x}-\vec{v}_At_A,\vec{y}^\prime)
     (\vec{a}\cdot\vec{S})_{m_\alpha m_{\alpha}^{\prime}}
     (\vec{b}\cdot\vec{S})_{m_\beta m_{\beta}^{\prime}}.
   \end{split}
   \label{cab_korel_correl_function_XXX}
 \end{equation}
 When $t_A=t_B=t$ the correlation function
 \eqref{cab_korel_correl_function_XXX} takes the form
 \begin{equation} 
 {\mathcal{C}}^{\alpha\beta}_{\psi}(\vec{a},\vec{b}) = 
 \int_A d^3 \vec{x} \int_B d^3 \vec{y} 
 \sum_{\substack{m_\alpha,m_\beta\\
 m_{\alpha}^{\prime},m_{\beta}^{\prime}}} 
 \psi^{\star}_{m_{\alpha}^{\prime}m_{\beta}^{\prime}} 
 (\vec{x}-\vec{v}_At,\vec{y}-\vec{v}_Bt) \,
 \psi_{m_{\alpha}m_{\beta}}
 (\vec{x}-\vec{v}_At,\vec{y}-\vec{v}_Bt)
 (\vec{a}\cdot\vec{S})_{m_\alpha m_{\alpha}^{\prime}}\,
 (\vec{b}\cdot\vec{S})_{m_\beta m_{\beta}^{\prime}}.
 \label{cab_korel_equal_time_diff_no_dest}
 \end{equation}

 \subsubsection*{\bfseries The case $s=\frac12$}
 
 Now let us apply the formula
 (\ref{cab_korel_equal_time_diff_no_dest}) to the case of the lowest
 nontrivial spin $s=\frac12$.  The state $|\psi\rangle$ can be a triplet or a
 singlet.
 
 \textit{Singlet state}.  For the singlet state we have
 \begin{equation}
   \psi_{m_\alpha m_\beta}(\vec{x},\vec{y})= -\psi_{m_\beta
     m_\alpha}(\vec{x},\vec{y})
 \end{equation}
 and then from Eq.~\eqref{cab_korel_equal_time_diff_no_dest} we
 receive
 \begin{equation} 
   {\mathcal{C}}^{\alpha\beta}_{\psi_{\text{\scriptsize singlet}}}
   (\vec{a},\vec{b})  = 
   - \frac{1}{2} \cos{(\theta_{ab})}
   \int_A d^3 \vec{x} \int_B d^3 \vec{y}  
   \left| \psi_{\text{\footnotesize singlet}}
     (\vec{x}-\vec{v}_At,\vec{y}-\vec{v}_Bt) 
   \right|^2,
   \label{cab_korel_rozne_singlet}
 \end{equation}
 where $\psi_{\text{\footnotesize singlet}}(\vec{x},\vec{y})\equiv
 \psi_{1/2,-1/2}(\vec{x},\vec{y})$, $\theta_{ab}$ denotes an angle between
 vectors $\vec{a}$ and $\vec{b}$, and the normalization yields
 \begin{equation}
   \iint d^3 \vec{x}\, d^3 \vec{y} 
   |\psi_{\text{\footnotesize singlet}}(\vec{x},\vec{y})|^2
   =\tfrac{1}{2}. 
 \end{equation}
 Thus the correlation function depends on the vectors $\vec{a}$ and
 $\vec{b}$ in the standard way, i.e., it behaves like a cosine of an
 angle between vectors $\vec{a}$ and $\vec{b}$.  The only difference
 is the presence of the term $\int_A d^3 \vec{x} \int_B d^3 \vec{y} \left|
   \psi_{\text{\footnotesize singlet}}
   (\vec{x}-\vec{v}_At,\vec{y}-\vec{v}_Bt) \right|^2$, which
 influences the intensity of the correlations.
 
 \textit{Triplet state}.  In the triplet case we have
 \begin{equation}
   \psi_{m_\alpha m_\beta}(\vec{x},\vec{y})= \psi_{m_\beta
     m_\alpha}(\vec{x},\vec{y})
 \end{equation}
 and the correlation function has a rather complicated form,
 \begin{equation}
   \begin{split}
     {\mathcal{C}}^{\alpha\beta}_{\psi_{\mbox{\scriptsize triplet}}}
     (\vec{a},\vec{b}) = & 
     \tfrac{1}{4} 
     \int_A d^3 \vec{x} \int_B d^3 \vec{y}
     \{(|\psi_{++}|^2+|\psi_{--}|^2) \cos{\theta_a}\cos{\theta_b}
     + (\psi^{\star}_{++}\psi_{--}e^{i(\varphi_a+\varphi_b)}+\psi^{\star}_{--}\psi_{++}e^{-i(\varphi_a+\varphi_b)}) \sin{\theta_a}\sin{\theta_b}\\
     &\quad
     +(\psi^{\star}_{++}\psi_{+-}-\psi^{\star}_{+-}\psi_{--}) (\cos{\theta_a}\sin{\theta_b}e^{i\varphi_b}+\sin{\theta_a}\cos{\theta_b}e^{i\varphi_a})
     +(\psi^{\star}_{+-}\psi_{++}-\psi^{\star}_{--}\psi_{+-}) (\cos{\theta_a}\sin{\theta_b}e^{-i\varphi_b}\\
     &\quad+\sin{\theta_a}\cos{\theta_b}e^{-i\varphi_a})
     -2 \psi^{\star}_{+-}\psi_{+-} [\cos{\theta_a}\cos{\theta_b}-\sin{\theta_a}\sin{\theta_b}\cos{(\varphi_a-\varphi_b)}]\},
   \end{split}
   \label{cab_korel_triplet_correl_function}
 \end{equation}
\end{widetext}
where we used the following notation:
\begin{gather}
  \psi_{++} \equiv \psi_{1/2,1/2}(\vec{x}-\vec{v}_At,\vec{y}-\vec{v}_Bt),\\
  \psi_{+-} \equiv \psi_{1/2,-1/2}(\vec{x}-\vec{v}_At,\vec{y}-\vec{v}_Bt),\\
  \psi_{--} \equiv \psi_{-1/2,-1/2}(\vec{x}-\vec{v}_At,\vec{y}-\vec{v}_Bt),\\
  \vec{a} = (\cos{\varphi_a}\sin{\theta_a}, \sin{\varphi_a}\sin{\theta_a}, \cos{\theta_a}),\\
  \vec{b} = (\cos{\varphi_b}\sin{\theta_b}, \sin{\varphi_b}\sin{\theta_b}, \cos{\theta_b}),
\end{gather}
and the normalization yields
\begin{multline}
  \iint d^3 \vec{x}\, d^3 \vec{y} \{|\psi_{1/2,1/2}(\vec{x},\vec{y})|^2+
  |\psi_{-1/2,-1/2}(\vec{x},\vec{y})|^2+\\
  2|\psi_{\frac{1}{2},-\frac{1}{2}}(\vec{x},\vec{y})|^2\} = 1.
\end{multline} 
The triplet correlation function given by
Eq.~\eqref{cab_korel_triplet_correl_function} depends on velocities of
frames in a more non-trivial way than in the singlet case.  Moreover,
it is evident that the dependence on frame velocities in fact reduces
to the dependence on the relative velocity of the observers.

\subsection{Identical particles}

Now we calculate the correlation function in the same setting as in
the previous subsection but in the case of two identical particles. We
denote the one-particle Hilbert space by ${\mathcal{H}}$.  For our
purpose it is convenient to use in ${\mathcal{H}}$ the basis
$\{|\vec{x},\vec{n},\lambda\rangle\}$, where, as previously, $|\vec{x},\vec{n},\lambda\rangle$
represents the particle localized at $\vec{x}$ and with spin component
along the direction $\vec{n}$ equal to $\lambda$ [see also the Appendix,
Eqs.~\eqref{cab_korel_vectors_xnlambda_eq1} and
\eqref{cab_korel_vectors_xnlambda_eq2}].  Since the particles are
identical, the state vectors of the two-particle system constitute the
symmetrical or antisymmetrical subspace of
${\mathcal{H}}\otimes{\mathcal{H}}$ and the observables are represented by
symmetrical operators.  Thus an observable that measures the spin
component along direction $\Vec{n}$ inside the region $\Omega$ has the form
\begin{equation}
  \Delta_{\Omega,\vec{n}}^{s}=\Lambda_{\Omega,\vec{n}}^{s}
  \otimes I+I\otimes\Lambda_{\Omega,\vec{n}}^{s}=
  \sum_{\lambda=-s}^{s}\lambda(
  \Pi_{\Omega,\vec{n}}^{s,\lambda}\otimes
  I+I\otimes\Pi_{\Omega,\vec{n}}^{s,\lambda}),
  \label{cab_korel_delta}
\end{equation}
where, as in the previous subsection,
\begin{gather}
  \Pi_{\Omega,\vec{n}}^{s,\lambda}\equiv\int_{\Omega}d^3 \vec{x}\,
  |\vec{x},\vec{n},\lambda\rangle\langle\vec{x},\vec{n},\lambda|,
  \label{cab_korel_Pi_Omega_n_s}\\
  \Lambda_{\Omega,\vec{n}}^{s}\equiv
  \sum_{\lambda=-s}^{s}\lambda\left(\int_{\Omega}d^3 \vec{x}\,
    |\vec{x},\vec{n},\lambda\rangle\langle\vec{x},\vec{n},\lambda|\right)
  =\sum_{\lambda=-s}^{s}\lambda\,\,
  \Pi_{\Omega,\vec{n}}^{s,\lambda},
\end{gather}
and $I$ denotes the identity. When we apply the observable
\eqref{cab_korel_delta}, we do not know how many particles are inside
$\Omega$.  Thus $\Delta_{\Omega,\vec{n}}^{s}$ measures the component of the total
spin of all particles inside $\Omega$.  To justify the above statement, we
consider the particle number operator $\Pi_{\Omega}^{s}\otimes I+I\otimes \Pi_{\Omega}^{s}$,
where
\begin{equation}
  \Pi_{\Omega}^{s}\equiv \sum_{\lambda=-s}^{s}
  \int_{\Omega}d^3 \vec{x}\,
  |\vec{x},\vec{n},\lambda\rangle\langle\vec{x},\vec{n},\lambda|=
  \sum_{\lambda=-s}^{s}
  \Pi_{\Omega,\vec{n}}^{s,\lambda}, 
\end{equation}
which discriminates how many particles are inside $\Omega$.  The spectral
decomposition of $\Pi_{\Omega}^{s}\otimes I+I\otimes \Pi_{\Omega}^{s}$ reads
\begin{equation}
  \Pi_{\Omega}^{s}\otimes
  I+I\otimes \Pi_{\Omega}^{s}=2 \Pi_{\Omega}^{(2)}+
  1 \Pi_{\Omega}^{(1)}+0 \Pi_{\Omega}^{(0)},
  \label{cab_korel_spectral_Pi_I_plus_I_Pi}
\end{equation}
where $\Pi_{\Omega}^{(2)}$, $\Pi_{\Omega}^{(1)}$, $\Pi_{\Omega}^{(0)}$ are projectors on
mutually orthogonal subspaces and their explicit form is the
following:
\begin{gather}
  \Pi_{\Omega}^{(2)} =\Pi_{\Omega}^{s}\otimes\Pi_{\Omega}^{s},
  \label{cab_korel_operator_Pi_(2)}\\ 
  \Pi_{\Omega}^{(1)} =\Pi_{\Omega}^{s}\otimes I +
  I\otimes\Pi_{\Omega}^{s} -2 \Pi_{\Omega}^{s}\otimes\Pi_{\Omega}^{s},
  \label{cab_korel_operator_Pi_(1)}\\ 
  \Pi_{\Omega}^{(0)} =I\otimes I- \Pi_{\Omega}^{s}\otimes I -
  I\otimes\Pi_{\Omega}^{s} +
  \Pi_{\Omega}^{s}\otimes\Pi_{\Omega}^{s}.
  \label{cab_korel_operator_Pi_(0)} 
\end{gather}
Therefore, as a result of measurement of $\Pi_{\Omega}^{s}\otimes I+I\otimes \Pi_{\Omega}^{s}$,
we receive one of the following outcomes (i)~there is no particle
inside $\Omega$; (ii)~there is one particle inside $\Omega$; and (iii)~there are
two particles inside $\Omega$.

In the sequel we will restrict ourselves to the simplest case
$s=\frac12$.  Let us denote for simplicity
\begin{equation}
  \Pi_{\Omega,\vec{n}}^{+}\equiv
  \Pi_{\Omega,\vec{n}}^{1/2,1/2}, 
  \qquad 
  \Pi_{\Omega,\vec{n}}^{-}\equiv
  \Pi_{\Omega,\vec{n}}^{1/2,-1/2}. 
  \label{cab_korel_Pi_plus_Pi_minus}
\end{equation}
The spectral decomposition of $\Delta_{\Omega,\vec{n}}\equiv\Delta_{\Omega,\vec{n}}^{1/2}$ has
the form
\begin{multline}
  \Delta_{\Omega,\vec{n}}= 
  \tfrac{1}{2} \Pi_{\Omega,\vec{n}}^{(1,+)}
  -\tfrac{1}{2} \Pi_{\Omega,\vec{n}}^{(1,-)} +
  1 \Pi_{\Omega,\vec{n}}^{(2,1)}
  -1 \Pi_{\Omega,\vec{n}}^{(2,-1)}\\
 +0 \Pi_{\Omega,\vec{n}}^{(2,0)} + 0 \Pi_{\Omega}^{(0,0)},
\end{multline}
where
\begin{gather}
  \begin{split}
    \Pi_{\Omega,\vec{n}}^{(1,+)}  &= \Pi_{\Omega,\vec{n}}^{+}\otimes I +
    I\otimes\Pi_{\Omega,\vec{n}}^{+} - 2  \Pi_{\Omega,\vec{n}}^{+}\otimes\Pi_{\Omega,\vec{n}}^{+}\\ 
    &\quad -  \Pi_{\Omega,\vec{n}}^{+}\otimes\Pi_{\Omega,\vec{n}}^{-}-
    \Pi_{\Omega,\vec{n}}^{-}\otimes\Pi_{\Omega,\vec{n}}^{+},
  \end{split}
  \label{cab_korel_projektory_N_lambda_1}\\
  \begin{split}
    \Pi_{\Omega,\vec{n}}^{(1,-)} &= \Pi_{\Omega,\vec{n}}^{-}\otimes I +
    I\otimes\Pi_{\Omega,\vec{n}}^{-} - 2 \Pi_{\Omega,\vec{n}}^{-}\otimes\Pi_{\Omega,\vec{n}}^{-}\\ 
    &\quad - \Pi_{\Omega,\vec{n}}^{+}\otimes\Pi_{\Omega,\vec{n}}^{-}-
    \Pi_{\Omega,\vec{n}}^{-}\otimes\Pi_{\Omega,\vec{n}}^{+},
  \end{split}
  \label{cab_korel_projektory_N_lambda_1_1}\\
  \Pi_{\Omega,\vec{n}}^{(2,1)} = \Pi_{\Omega,\vec{n}}^{+} \otimes \Pi_{\Omega,\vec{n}}^{+},\\
  \Pi_{\Omega,\vec{n}}^{(2,-1)} = \Pi_{\Omega,\vec{n}}^{-} \otimes \Pi_{\Omega,\vec{n}}^{-},\\
  \Pi_{\Omega,\vec{n}}^{(2,0)} = \Pi_{\Omega,\vec{n}}^{+} \otimes \Pi_{\Omega,\vec{n}}^{-} + \Pi_{\Omega,\vec{n}}^{-} \otimes \Pi_{\Omega,\vec{n}}^{+},\\
  \begin{split}
    \Pi_{\Omega,\Vec{n}}^{(0,0)} &= I \otimes I - \Pi_{\Omega,\vec{n}}^{+} \otimes I - \Pi_{\Omega,\vec{n}}^{-} \otimes I
    - I \otimes \Pi_{\Omega,\vec{n}}^{+} - I \otimes \Pi_{\Omega,\vec{n}}^{-}\\
    &\quad + \Pi_{\Omega,\vec{n}}^{+} \otimes \Pi_{\Omega,\vec{n}}^{+} + \Pi_{\Omega,\vec{n}}^{-}\otimes\Pi_{\Omega,\vec{n}}^{-} 
    + \Pi_{\Omega,\vec{n}}^{+}\otimes\Pi_{\Omega,\vec{n}}^{-}\\ 
    &\quad +  \Pi_{\Omega,\vec{n}}^{-}\otimes\Pi_{\Omega,\vec{n}}^{+}
  \end{split}
  \label{cab_korel_projektory_N_lambda_2}
\end{gather}
are projectors on mutually orthogonal subspaces and one can easily
check that
\begin{gather}
  \Pi_{\Omega,\vec{n}}^{(2)} = \Pi_{\Omega,\vec{n}}^{(2,1)}+
  \Pi_{\Omega,\vec{n}}^{(2,-1)}+ \Pi_{\Omega,\vec{n}}^{(2,0)},\\
  \Pi_{\Omega,\vec{n}}^{(1)} = \Pi_{\Omega,\vec{n}}^{(1,+)}+\Pi_{\Omega,\vec{n}}^{(1,-)},\\
  \Pi_{\Omega,\vec{n}}^{(0)} = \Pi_{\Omega,\vec{n}}^{(0,0)}
\end{gather}
[cf.\ 
Eqs.~\eqref{cab_korel_operator_Pi_(2)}--\eqref{cab_korel_operator_Pi_(0)}].
Thus we can see that the observable $\Delta_{\Omega,\vec{n}}$ really measures
the component of the total spin of all the particles inside $\Omega$.

Now we are prepared to calculate quantum correlations in the case of
two identical particles.  As previously, we assume that a two-particle
state $\rho(t_A)$ is prepared at time $t_A$ in a certain inertial frame
of reference ${\mathcal{O}}$.  Moreover, the two observers,
${\mathcal{A}}$ and ${\mathcal{B}}$, move with constant velocities
with respect to ${\mathcal{O}}$.  We denote the velocity of the frame
${\mathcal{O}}$ with respect to ${\mathcal{A}}$ and ${\mathcal{B}}$ by
$\vec{v}_A$ and $\vec{v}_B$, respectively.  The observer
${\mathcal{A}}$ measures at time $t_A$ the observable $\Delta_{A,\Vec{a}}$,
and as a result of measurement with selection he receives the value
$\lambda$.  Here $A$ denotes some bounded region in ${\mathbb{R}}^3$ and
$\vec{a}$ denotes a unit vector.  Next, at time $t_B \geq t_A$ the
observer ${\mathcal{B}}$ measures $\Delta_{B,\Vec{b}}$ and receives $\lambda^\prime$,
where similarly $B\subset{\mathbb{R}}^3$ and $\vec{b}$ is a unit vector.  We
can write
\begin{gather}
  \Delta_{A,\Vec{a}} = \sum_{N=0}^{2}\left(\sum_{\lambda_{N}} \lambda_{N}
    \Pi_{A,\Vec{a}}^{(N,\lambda_{N})}\right), \label{cab_korel_Delta_A}\\
  \Delta_{B,\Vec{b}} =  \sum_{N=0}^{2}\left(\sum_{\lambda_{N}} \lambda_{N}
    \Pi_{B,\Vec{b}}^{(N,\lambda_{N})}\right). \label{cab_korel_Delta_B}
\end{gather}

For the explicit form of $\Pi_{A,\Vec{a}}^{(N,\lambda_{N})}$ and
$\Pi_{B,\Vec{b}}^{(N,\lambda_{N})}$, see Eqs.~\eqref{cab_korel_projektory_N_lambda_1}--%
\eqref{cab_korel_projektory_N_lambda_2}.  In
Eqs.~\eqref{cab_korel_Delta_A}--\eqref{cab_korel_Delta_B}, $\lambda_N$ is an
eigenvalue of $\Delta_{A,\Vec{a}}$, $\Delta_{B,\Vec{b}}$ and the projector
$\Pi_{A,\Vec{a}}^{(N,\lambda_{N})}$ corresponds to the situation in which
inside the region $A$ there are $N$ particles and the total spin
component along the direction $\Vec{a}$ of all these particles is
equal to $\lambda_N$.

As in the previous section, let $p(\lambda,\lambda^\prime)$ denote the probability that
${\mathcal{A}}$ receives the value $\lambda$ and ${\mathcal{B}}$ the value
$\lambda^\prime$.  We define the correlation function by the formula
\begin{equation}
  {\cal C}(\vec{a},\vec{b})=\sum_{\lambda,\lambda^\prime} \lambda \lambda^\prime p(\lambda,\lambda^\prime).
  \label{cab_korel_funkcja_identyczne_bez_an}
\end{equation}
This function differs from zero when each observer registers one
particle or when one observer registers two particles and the second
one registers one or two particles.  If one of them registers no
particle, then the corresponding $\lambda$ is equal to zero.  We take into
account only the case in which each observer registers one particle.

Performing similar steps as in the case of distinguishable particles
and taking into account Eqs.~\eqref{cab_korel_Delta_A} and
\eqref{cab_korel_Delta_B}, we get
\begin{widetext}
  \begin{equation}
    p(\lambda_1,\lambda_{1}^{\prime}) = \mathrm{Tr} \{[{\mathcal{U}}^{\dagger}_{t_A}(\vec{v}_A)\,
    \Pi_{A,\vec{a}}^{(1,\lambda_1)}\, {\mathcal{U}}_{t_A}(\vec{v}_A)]
    \rho(t_A) [{\mathcal{U}}^{\dagger}_{t_A}(\vec{v}_A)\,
    \Pi_{A,\vec{a}}^{(1,\lambda_1)}\, {\mathcal{U}}_{t_A}(\vec{v}_A)]
    [{\mathcal{U}}(t_B-t_A)\, {\mathcal{U}}^{\dagger}_{t_B}(\vec{v}_B)\,
    \Pi_{B,\vec{b}}^{(1,\lambda_{1}^{\prime})}\, {\mathcal{U}}_{t_B}(\vec{v}_B)\,
    {\mathcal{U}}^\dagger(t_B-t_A)]\},
    \label{cab_korel_prawdop_pary_identyczne_bez_anihilacji}
  \end{equation}
  where, as previously
  \begin{equation}
    {\mathcal{U}}_t(\vec{v})=U_t(\vec{v})\otimes U_t(\vec{v}),\quad 
    {\mathcal{U}}(t)=U(t)\otimes U(t)
  \end{equation}
  and $U_t(\Vec{v})$, $U(t)$ are defined in the Appendix; see
  Eqs.~\eqref{cab_korel_app_unitary_representants},
  \eqref{cab_korel_app_action_momentum_eq4},
  \eqref{cab_korel_app_action_position_eq4},
  Eqs.~\eqref{cab_korel_app_action_position_n_eq4}.  Inserting
  Eq.~\eqref{cab_korel_prawdop_pary_identyczne_bez_anihilacji} into
  Eq.~\eqref{cab_korel_funkcja_identyczne_bez_an}, we arrive at the
  following formula for the correlation function:
  \begin{equation}
    \begin{split}
    {\cal C}(\vec{a},\vec{b}) &=
    \mathrm{Tr}\left\{\sum_{\lambda_1}
      [{\mathcal{U}}^{\dagger}_{t_A}(\vec{v}_A)\, \Pi_{A,\Vec{a}}^{(1,\lambda_1)}\, 
      {\mathcal{U}}_{t_A}(\vec{v}_A)] \rho(t_A)
      [{\mathcal{U}}^{\dagger}_{t_A}(\vec{v}_A)\, \Pi_{A,\Vec{a}}^{(1,\lambda_1)}\, 
      {\mathcal{U}}_{t_A}(\vec{v}_A)] [{\mathcal{U}}(t_B-t_A)\,
      {\mathcal{U}}^{\dagger}_{t_B}(\vec{v}_B)\, 
      \Delta_{B,\Vec{b}}\, {\mathcal{U}}_{t_B}(\vec{v}_B)\,
      {\mathcal{U}}^\dagger(t_B-t_A)] \right.\\
    &\quad\times \left.\vphantom{\sum_{\lambda_1}}
      [{\mathcal{U}}^{\dagger}_{t_A}(\vec{v}_A)\, \Delta_{A,\vec{a}}\,
      {\mathcal{U}}_{t_A}(\vec{v}_A)] \right\}.
  \end{split}
  \label{cab_korel_f_kor_ident_bez_an_rozne_czasy_rozne_uklady}
 \end{equation}
\end{widetext}

Let us consider the simplest case in which both the observers rest
with respect to ${\mathcal{O}}$ and the measurements are performed at
the same time.  It means that we put in
Eq.~\eqref{cab_korel_f_kor_ident_bez_an_rozne_czasy_rozne_uklady}
\begin{equation}
  \vec{v}_A=\vec{0},\quad \vec{v}_B=\vec{0},\quad t_A=t_B=t.
\end{equation}
The result is the following
\begin{equation}
  {\cal C}(\vec{a},\vec{b}) = \sum_{\lambda_1}
  \mathrm{Tr}\{\Pi_{A,\Vec{a}}^{(1,\lambda_1)}\, \rho(t)\, \Pi_{A,\vec{a}}^{(1,\lambda_1)}\,
  \Delta_{B,\vec{b}}\,\Delta_{A,\vec{a}}\}. 
  \label{f_kor_ident_bez_an_ten_sam_uklad_ten_sam_czas}
\end{equation}
Furthermore, if we assume that the regions $A$ and $B$ are disjoint
(as it happens in the real EPR-type experiments) we find from
\eqref{cab_korel_delta},\ref{cab_korel_Pi_Omega_n_s},
\eqref{cab_korel_Pi_plus_Pi_minus},
\eqref{cab_korel_projektory_N_lambda_1}, and
\eqref{cab_korel_projektory_N_lambda_1_1} that $\Delta_{B,\Vec{b}}$ and
$\Pi_{A,\Vec{a}}^{(1,\lambda_1)}$ commute.  Thus changing the order in
Eq.~\eqref{f_kor_ident_bez_an_ten_sam_uklad_ten_sam_czas} and using
Eq.~\ref{cab_korel_Delta_A} we get
\begin{equation}
  {\cal C}(\vec{a},\vec{b}) = 
  \mathrm{Tr}\{\rho(t)\,  \Delta_{A,\vec{a}}\, \Delta_{B,\vec{b}}\}. 
  \label{f_kor_ident_bez_an_ten_sam_uklad_ten_sam_czas_eq_2}
\end{equation}

\section{\label{cab_korel_sec_conclusions}Conclusions}

In this paper, we have presented the detailed calculation of the spin
correlation functions in the EPR-type experiments.  In opposition to
the standard approach, we considered the space part of the wave
function and the relative motion of the observers.  We also took into
account the fact that every measurement of the spin component is
connected with the simultaneous localization of the particle inside
the detector.  Thus we assumed that observers perform measurements in
some bounded regions of space.  We performed our calculations in the
framework of nonrelativistic quantum mechanics.  There were two main
reasons for this.  First, to take into account the localization in the
regions of the detectors, we need the well-defined notion of the
localization (in the standard formulation of relativistic quantum
mechanics, the notion of localization of the particles is ill-defined,
however see \cite{cab_korel_CR1999}).  Secondly, to the best of our
knowledge, any systematic review of EPR correlations in the framework
of nonrelativistic quantum mechanics in the general case mentioned
above does not exist in the literature.  Moreover, we considered
separately the case of identical and distinguishable particles.  In
both cases, we found the general formula for the correlation function
under the assumption that the initial two-particle state is arbitrary.
In addition, we also considered some important special cases.

In the most interesting case of the singlet state of two
spin-$\frac12$ particles, we determined, as one might expect, that the
correlation function depends on the vectors $\vec{a}$ and $\vec{b}$ in
the standard way, i.e., it behaves like a cosine of an angle between
vectors $\vec{a}$ and $\vec{b}$.  The only difference is the presence
of the term $\int_A d^3 \vec{x} \int_B d^3 \vec{y} \left|
  \psi_{\text{\footnotesize singlet}}
  (\vec{x}-\vec{v}_At,\vec{y}-\vec{v}_Bt) \right|^2$, which influences
the intensity of the correlations.  Note that the low velocity limit
of the relativistic correlation function for the singlet state of two
spin-$\frac12$ particles derived in
\cite{cab_korel_Czachor1997_1,cab_korel_Czachor1997_2} and in
\cite{cab_korel_RS2002} differs from our results.  The correlation
function obtained in
\cite{cab_korel_Czachor1997_1,cab_korel_Czachor1997_2} depends on the
state of motion of the particles, while the function derived in
\cite{cab_korel_RS2002} contains a correction of second order in
velocities to our nonrelativistic formula.  However, in this second
case, when both measurements are performed in the same inertial frame,
the limiting correlation function coincides with our results.

\appendix*
\section{Galilean group} 

To establish notation and conventions, we summarize here the main
facts concerning the Galilean group and its unitary representations.
Let $\mathcal{H}$ be the one-particle Hilbert space of states. In this
space, the following basic observables exist: $\Hat{\Vec{X}}$
(position), $\Hat{\Vec{P}}$ (momentum), $\Hat{\Vec{S}}$ (spin).  They
fulfil the following relations:
\begin{gather}
 [\hat{X_i},\hat{X}_j]=0, \quad
 [\hat{P_i},\hat{P}_j]=0, \label{cab_korel_observables_1}\\
 [\hat{X}_i,\hat{P}_j]=i\delta_{ij}, \quad
 [\hat{X}_i,\hat{S}_j]=0,\\ 
 [\hat{P}_i,\hat{S}_j]=0, \quad
 [\hat{S}_i,\hat{S}_j]=i\varepsilon_{ijk}\hat{S}_k. 
 \label{cab_korel_observables_3}
\end{gather}
The Galilean group and its algebra act in the space $\mathcal{H}$.  As
is well known, classical Galilean transformations have the following
form
\begin{gather}
  \vec{x}^\prime = R\vec{x}+\vec{a}-\vec{v}t,\\
  t^\prime = t+\tau,
\end{gather}
where $\vec{v}$ denotes the velocity of the frame $(\vec{x}^\prime,t^\prime)$
with respect to the frame $(\vec{x},t)$ and we adopt the passive point
of view.  In the Hilbert, space the rotation $R$ is generated by the
total angular momentum $\Hat{\Vec{J}}$, the translation $\vec{a}$ is
generated by the momentum $\Hat{\Vec{P}}$, the time translation is
generated by the Hamiltonian $\hat{H}$ and Galilean boost is generated
by $\Hat{\Vec{G}}$.  The basic commutation relations of the Galilean
algebra (in fact its central extension) read
\begin{gather}
  [\hat{P}_i,\hat{P}_j] =0, \quad
  [\hat{P}_i,\hat{G}_j] =i\delta_{ij}MI \\
  [\hat{G}_i,\hat{G}_j] =0, \quad
  [\hat{H},\hat{P}_i] =0,\\
  [\hat{J}_i,\hat{J}_j] =i\varepsilon_{ijk}\hat{J}_k, \quad
  [\hat{H},\hat{G}_i] =i\hat{P}_i,\\
  [\hat{J}_i,\hat{P}_j] =i\varepsilon_{ijk}\hat{P}_k, \quad
  [\hat{H},\hat{J}_i] =0,\\
  [\hat{J}_i,\hat{G}_j] =i\varepsilon_{ijk}\hat{G}_k, \quad
  i,j,k=1,2,3,
\end{gather}
where $M$ is the mass of the system.  All generators of Galilean
transformations can be expressed by observables
\eqref{cab_korel_observables_1}--\eqref{cab_korel_observables_3}.  We
have
\begin{equation}
  \Hat{\Vec{J}}=\Hat{\Vec{S}}+\Hat{\Vec{X}}\times\Hat{\Vec{P}},\quad
  \Hat{\Vec{G}}=t\Hat{\Vec{P}}-M\Hat{\Vec{X}}.
\end{equation}
Moreover in the case of a free particle $\hat{H}=\Hat{\Vec{P}}^2/2M$.

Irreducible unitary ray representations of the Galilean group are
determined by two numbers: the eigenvalue of $\Hat{\Vec{S}}^2$, which
has the form $s(s+1)$, where $s$ is an integer or a half-integer and
the non-negative real constant $M$.  In the sequel we will use the
Schr{\"o}dinger picture.  Since $\Hat{\Vec{G}}$ depends on time explicitly,
this dependence remains also in Sch{\"o}dinger picture.  In momentum
representation, we will denote the basis vectors of the carrier space
of the determined irreducible unitary representation of the Galilean
group (the space of states of the system at time $t$) by
$|\vec{k},m\rangle_t$, where $\vec{k}$ is an eigenvalue of momentum operator
$\Hat{\Vec{P}}$, and $m$ is the spin component along the $z$ axis.  We
will denote elements of the unitary representation of the Galilean
group as follows:
\begin{gather}
  U(\vec{a})=e^{i\vec{a}\cdot\Hat{\Vec{P}}}, \quad
  U(\vec{v})=e^{i\vec{v}\cdot\Hat{\Vec{G}}}, \nonumber\\
  U(R)=e^{i\boldsymbol{\varphi}\cdot\Hat{\Vec{J}}}, \quad
  U(\tau)=e^{i\tau\hat{H}},
  \label{cab_korel_app_unitary_representants}
\end{gather}
where $\vec{a}$, $\vec{v}$, $\boldsymbol{\varphi}$, and $\tau$ are parameters
corresponding to pure translations, Galilean boosts, rotations, and
time translation, respectively.  The action of operators
\eqref{cab_korel_app_unitary_representants} on basis vectors is the
following:
\begin{gather}
  U(\vec{a})\,|\vec{k},m\rangle_t =
  e^{i\vec{a}\cdot\vec{k}}\,|\vec{k},m\rangle_t,  
  \label{cab_korel_app_action_momentum_eq2}\\
  U(R)\,|\vec{k},m\rangle_t  = {\mathcal{D}}^s(R)_{m^\prime m}\,
  |R\vec{k},m^\prime\rangle_t,
  \label{cab_korel_app_action_momentum_eq3}\\
  U_t(\vec{v})\,|\vec{k},m\rangle_t = 
  e^{it[\vec{v}\cdot\vec{k}-(M v^2/2)]}\, |\vec{k}-M\vec{v},m\rangle_t,
  \label{cab_korel_app_action_momentum_eq4}
\end{gather}
where ${\mathcal{D}}^s(R)$ is the spin $s$ irreducible unitary
representation of $\mathrm{SU}(2)$.  Moreover we, assume that
\begin{equation}
  e^{i\boldsymbol{\alpha}\cdot\Hat{\Vec{X}}}\,|\vec{k},m\rangle_t =
  |\vec{k}+\boldsymbol{\alpha},m\rangle_t.
  \label{cab_korel_app_action_momentum_eq1}\\
\end{equation}
The phase factors in Eqs.~\eqref{cab_korel_app_action_momentum_eq4}
and \eqref{cab_korel_app_action_momentum_eq1} determine the following
form of the vector $|\vec{k},m\rangle_{t+\tau}$:
\begin{equation}
  |\vec{k},m\rangle_{t+\tau}=e^{-i\tau(\vec{k}^2/2M)}
  |\vec{k},m\rangle_t.
\end{equation}
We denote eigenvectors of the position operator $\Hat{\Vec{X}}$ by
$|\vec{x},m\rangle_t$.  We have
\begin{gather}
  \Hat{\Vec{X}}|\vec{x},m\rangle_t=\vec{x}|\vec{x},m\rangle_t,\\
  |\vec{x},m\rangle_t=\frac{1}{(2\pi)^{3/2}}\int d^3 \vec{k}\,
  e^{-i\vec{k}\cdot\vec{x}}\, |\vec{k},m\rangle_t. \label{cab_korel_Fourier} 
\end{gather}
The Galilean group and operator $e^{i\boldsymbol{\alpha}\cdot\Hat{\Vec{X}}}$
act on vectors $|\vec{x},m\rangle_t$ as follows:
\begin{gather}
  U(\vec{a})\,|\vec{x},m\rangle_t = |\vec{x}-\vec{a},m\rangle_t, \\ 
  U(R)\,|\vec{x},m\rangle_t = {\mathcal{D}}^s(R)_{m^\prime m}\,
  |R\vec{x},m^\prime\rangle_t,\\
  U_t(\vec{v})\,|\vec{x},m\rangle_t =
  e^{iM[(t\vec{v}^2/2)-\vec{v}\cdot\vec{x}]}\,
  |\vec{x}-t\vec{v},m\rangle_t,\\ 
  e^{i\boldsymbol{\alpha}\cdot\Hat{\Vec{X}}}\,|\vec{x},m\rangle_t = 
  e^{i\boldsymbol{\alpha}\cdot\vec{x}}\,|\vec{x},m\rangle_t. 
  \label{cab_korel_app_action_position_eq4}
\end{gather}
In discussion of EPR-like experiments, it is convinient to use a
position basis in which vectors are numbered by spin component along
an arbitrary axis (not necessarily the $z$ axis).  Thus let $\vec{n}$
be an arbitrary unit vector.  Observable $\vec{n}\cdot\Hat{\Vec{S}}$
measures spin component along an axis in direction $\vec{n}$.  Since
$\vec{n}\cdot\Hat{\Vec{S}}$ commutes with $\Hat{\Vec{X}}$, these
observables possess a common set of eigenvectors.  We denote them by
$|\vec{x},\vec{n},\lambda\rangle$,
\begin{gather}
  (\vec{n}\cdot\Hat{\Vec{S}})\,|\vec{x},\vec{n},\lambda\rangle  
  = (\vec{n}\cdot\boldsymbol{\Sigma})_{\sigma\lambda}
  |\vec{x},\vec{n},\sigma\rangle = 
  \lambda\,|\vec{x},\vec{n},\lambda\rangle,
  \label{cab_korel_vectors_xnlambda_eq1}\\
  \Hat{\Vec{X}}\,|\vec{x},\vec{n},\lambda\rangle =
  \vec{x}\,|\vec{x},\vec{n},\lambda\rangle, 
  \label{cab_korel_vectors_xnlambda_eq2}
\end{gather}
where values of $\lambda$ are the same as values of $m$ ($\lambda=-s,-s+1,\ldots, s$)
and $\boldsymbol{\Sigma}$ denotes the generators of the representation
${\mathcal D}^s$.  If we parametrize $\vec{n}$ explicitly as follows:
\begin{equation}
  \vec{n}=(\sin{\theta}\cos{\varphi}, \sin{\theta}\sin{\varphi}, \cos{\theta}),
\end{equation}
we get
\begin{equation}
  |\vec{x},\vec{n},\lambda\rangle= {\mathcal{D}}^s
  (e^{i\theta\boldsymbol{\omega}\cdot\boldsymbol{\Sigma}})_{\lambda^\prime\lambda}  
  |\vec{x},\lambda^\prime\rangle_t,
\end{equation}
where the vector $\boldsymbol{\omega}$ is orthogonal to $\vec{n}$ and
\begin{equation}
  \boldsymbol{\omega}=(\sin{\varphi},-\cos{\varphi},0).
\end{equation}
We define also the vectors $|\vec{k},\vec{n},\lambda\rangle$ as follows:
\begin{equation}
  |\vec{k},\vec{n},\lambda\rangle=\frac{1}{(2\pi)^{3/2}}\int d^3 \vec{k}\,
  e^{i\vec{k}\cdot\vec{x}}\,
  |\vec{x},\vec{n},\lambda\rangle.
  \label{cab_korel_vectors_k_n_lambda} 
\end{equation}
One can also check that the following relations hold:
\begin{gather}
  \langle\vec{y},m|\vec{x},\vec{n},\lambda\rangle =
  \delta(\vec{x}-\vec{y}){\mathcal{D}}^{s}(e^{i\theta\boldsymbol{\omega}\cdot\boldsymbol{\Sigma}})_{m\lambda}, 
  \label{cab_korel_iloczyn_y_m_x_n_lambda}\\
  \langle\vec{p},m|\vec{x},\vec{n},\lambda\rangle =
  \frac{e^{-i\vec{p}\cdot\vec{x}}}{(2\pi)^{3/2}} 
  {\mathcal{D}}^{s}(e^{i\theta\boldsymbol{\omega}\cdot\boldsymbol{\Sigma}})_{m\lambda}. 
  \label{cab_korel_iloczyn_p_m_x_n_lambda}
\end{gather}
The Galilean group and operator $e^{i\boldsymbol{\alpha}\cdot\Hat{\Vec{X}}}$
act on the vectors $|\vec{x},\vec{n},\lambda\rangle$ as follows:
\begin{gather}
  U(\vec{a})\,|\vec{x},\vec{n},\lambda\rangle =
  |\vec{x}-\vec{a},\vec{n},\lambda\rangle, \\
  U(R)\,|\vec{x},\vec{n},\lambda\rangle = {\mathcal{D}}^s
  (R)_{\lambda^\prime\lambda}\, 
  |R\vec{x},R\vec{n},\lambda^\prime\rangle, \\
  U_t(\vec{v})\,|\vec{x},\vec{n},\lambda\rangle =
  e^{iM[(tv^2/2)-\vec{v}\cdot\vec{x}]}\,
  |\vec{x}-t\vec{v},\vec{n},\lambda\rangle,
  \label{cab_korel_app_action_position_n_eq4}\\
  e^{i\boldsymbol{\alpha}\cdot\Hat{\Vec{X}}}\,|\vec{x},\vec{n},\lambda\rangle =
  e^{i\boldsymbol{\alpha}\cdot\vec{x}}\,|\vec{x},\vec{n},\lambda\rangle.
\end{gather}
In the case $s=\frac{1}{2}$, we have
$\boldsymbol{\Sigma}=\frac{1}{2}\boldsymbol{\sigma}$, so
\begin{gather}
  |\vec{x},\vec{n},\tfrac{1}{2}\rangle =
  \cos(\theta/2)\,|\vec{x},\tfrac{1}{2}\rangle
  +e^{-i\varphi}\sin(\theta/2)\,|\vec{x},-\tfrac{1}{2}\rangle, \\
  |\vec{x},\vec{n},-\tfrac{1}{2}\rangle =
  -e^{i\varphi}\sin(\theta/2)\,|\vec{x},\tfrac{1}{2}\rangle
  +\cos(\theta/2)\,|\vec{x},-\tfrac{1}{2}\rangle.
\end{gather}


\end{document}